# Noise Reduction Technique for Raman Spectrum using Deep Learning Network

Liangrui Pan
Department of Electrical Engineering,
Faculty of Engineering
Prince of Songkla University
Thailand
lip141772@gmail.com

Pronthep Pipitsunthonsan
Department of Electrical Engineering,
Faculty of Engineering
Prince of Songkla University
Thailand
pronthep.pi@gmail.com

Peng Zhang
Department of Electrical Engineering,
Faculty of Engineering
Prince of Songkla University
Thailand
13195319733@163.com

Chalongrat Daengngam
Division of Physical Science
Faculty of Science
Prince of Songkla University
Thailand
chalongrat.d@psu.ac.th

Apidach Booranawong
Department of Electrical Engineering,
Faculty of Engineering
Prince of Songkla University
Thailand
apidet.boo@gmail.com

Mitchai Chongcheawchamnan
Department of Electrical Engineering,
Faculty of Engineering
Prince of Songkla University
Thailand
mitchai.c@psu.ac.th

*Abstract*—In a normal indoor environment, Raman spectrum encounters noise often conceal spectrum peak, leading to difficulty in spectrum intepretation. This paper proposes deep learning (DL) based noise reduction technique for Raman spectroscopy. The proposed DL network is developed with several training and test sets of noisy Raman spectrum. The proposed technique is applied to denoise and compare the performance with different wavelet noise reduction methods. Output signal-to-noise ratio (SNR), root-mean-square error (RMSE) and mean absolute percentage error (MAPE) are the performance evaluation index. It is shown that output SNR of the proposed noise reduction technology is 10.24 dB greater than that of the wavelet noise reduction method while the RMSE and the MAPE are 292.63 and 10.09, which are much better than the proposed technique.

## I. INTRODUCTION

Raman spectroscopy is a method widely used to describe the characteristics of materials, including industrial process control, planetary exploration, homeland security, life sciences, geological field investigations, and laboratory material research. In all these environments, it is necessary to obtain clear and effective Raman spectra as a prerequisite. However, the Raman spectrum signal will be mixed with various noises during the extraction process. To make Raman spectroscopy a useful tool for analyzing and detecting substances, it is necessary to reduce noise in the process of spectrum extraction. From the perspective of noise reduction of known signals, there are many methods to reduce noise, such as wavelet method, empirical mode decomposition method, etc. These methods have been able to reduce the noise in the signal very well. With its unique characteristics, Raman spectroscopy is different from other signals. There is a massive gap between peaks and valleys. Different peaks represent various properties of substances. Assuming that if noise masks part of the secondary peak information, subsequent experiments will be meaningless.

In the past decade, deep neural networks (DNNs) have been achieved for signal processing. It has been found that the performance of a deep neural network can be improved with the depth and structure of the neural network. The number of layers of neural networks has expanded to hundreds or even thousands of network layers [1], [2].

DNN usually incorporates a variety of known structures, including convolutional layers or LSTM units. Convolutional Neural Networks (CNN) and Recurrent Neural Networks (RNN) are the two main types of DNN architectures, which are widely used in various speech signal processing tasks [3]. CNN is characterized by being good at extracting position-invariant features, while RNN is characterized by being good at modelling units in sequence. Under normal circumstances, CNN is a layered architecture and RNN is a continuous structure. In RNN, the output of a neuron can directly affect itself in the next time period, that is, the input of the $i^{th}$ layer of neurons at time $m$, in addition to the output of the $(i - 1)^{th}$ layer of neurons at that time, it also includes itself at the output at time $m - 1$. At present, a use of CNNs to denoise noises in Raman spectra has yet to be studied [4].

Wavelet based denosing method was proposed for Raman spectroscopy and the performance is better than other methods [5]. However, there are also uncertain factors in this process, such as different wavelet, different frequency and different time will lead to the performance of noise reduction. The empirical mode decomposition (EMD) method does not need to analyze and study the unknown signal first, but directly decomposes it into the sum of several connotative modal components[6]. However, the intrinsic mode component has two constraints: (1) in the whole data segment, the number of extreme points and the number of zero-crossing points must be equal or the difference can not be more than one. (2) At any time, the average value of the upper envelope formed by the local maximum point and the lower envelope formed by the local minimum point is zero, that is, the upper and lower envelope lines are locally symmetric to the time axis.

To avoid the constraints of these methods, this paper proposes the deep learning (DL) network for noise reduction in Raman spectroscopy. Because the neural network is a nonlinear regression, it can not only satisfy the noise reduction of Raman spectrum signal, but also satisfy the noise reduction of other noisy signals. The structure of this paper is as follows: the second section will introduce the collection and collation of datasets. The third section will introduce DL model and evaluation index. The fourth section discusses the performance of the methods based on DL, CNN and wavelet. The final section is the conclusion.

## I. DATA COLLECTION AND PROCESSING

The preparation of Raman spectrum dataset will affect the implementation of all subsequent steps. First, we prepare noiseless Raman spectra of about 190 hazardous chemicals. It consists of one-dimensional signals with a length of 2,051 data points. The noise of Raman spectrum is mainly composed of fluorescence noise and Gaussian noise. The noise level will directly affect the quality of Raman spectroscopy. Since the amount of datasets is too small to meet the requirements of the hunger DL network, we added noises to the noiseless datasets. Firstly, baselines with different noise levels were added to the Raman spectrum. Then additive white Gaussian noise were added added up to the baseline noise contaminated Raman spectrum. White noise were addeded such that 0-80 dB SNR conditions were obtained. Shown in Table 1, 15,390 and 500 noisy Raman datasets were used for training and testing, respectively.

TABLE 1. Summary of training and testing environment

| Name | Value |
| --- | --- |
| Hazardous material | 190 |
| SNR | 0-80 dB |
| Training data | 15,390 |
| Testing data | 500 |

Baseline preprocessing is needed to ensure that most of the fluorescence noise has been removed before applying the denoising methods. In this paper, the most common method, airPLS algorithm, is chosen as the baseline correction method [7]. Fig. 1 shows the an example of the output spectrum after the correcction.

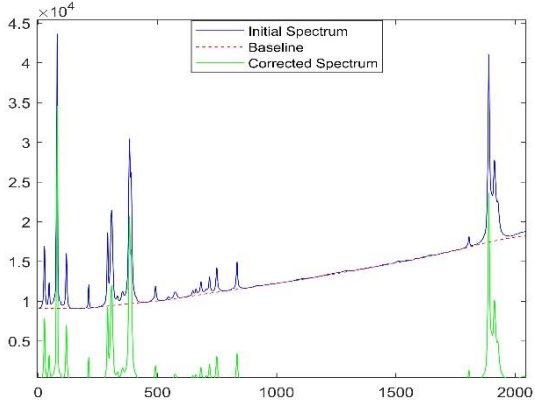

Fig. 1 Raman spectrum baseline correction process.

Subfigures shown in the left and right sides of Fig. 2 show the Raman spectra of two hazardous chemicals. Noiseless Raman spectrum is shown in Fig. 2 a) where Fig 2 b) and c) show Raman spectrum in 3 dB and 6 dB noise environment. Compared with the noiseless Raman spectrum signal, most of the peak information is submerged by noise in the secondary peak. In the main peak information, the noise increases the intensity of Raman spectrum, which makes the spectrum misleading in some cases.

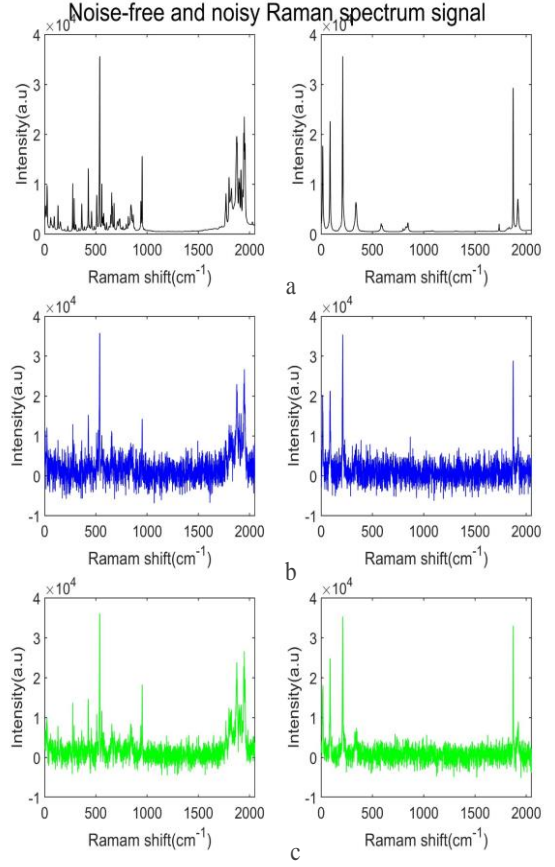

Fig. 2 Noiseless and noisy Raman spectrum signals (3 and 6 dB noise envionment)

## II. METHOD AND EVALUATION INDEX

### A. DL network

The DL model is implemented in MATLAB$^{TM}$ based on parallel convolutional neural network (CNNs). CNN neural network has been used to classify noise Raman spectra, but it has not been used to reconstruct Raman spectra. The current CNN model is obtained through experiments. It consists of 7 two-dimensional convolution layers. Each layer has 100 filters and the core size of each filter is 100×1. The first layer is a 2051×1×1 input layer, which is normalized by 'zerocenter'[8]. The number of samples per input Raman spectrum signal sequence (M) is 2,051. Each convolution layer has neurons connected to a part of the input feature or to the output of the previous layer. The step size (i.e. stripe) of the kernel is $[1 \times 1]$ and padding is introduced to equalize the output and input kernel size. After each convolution layer, there are 100-channel batch processing layer, rectifying linear unit (Relu) layer and the largest pooling layer with a stride of 2 and pooling size of 2. The continuity of these layers reduces the dimension of the input Raman spectrum sequence. Before the final regression output layer, the signal is regressed through a fully connected layer. DL model includes 14 convolution layers, 14 pooling layers, 14 maximum pooling layers, 14 Relu layers, one concat layer and 1 FC layer.

Fig. 3 describes the structure of the DL model. It can be assumed that the DL model will be able to learn the appropriate filter for noise reduction to recover the original

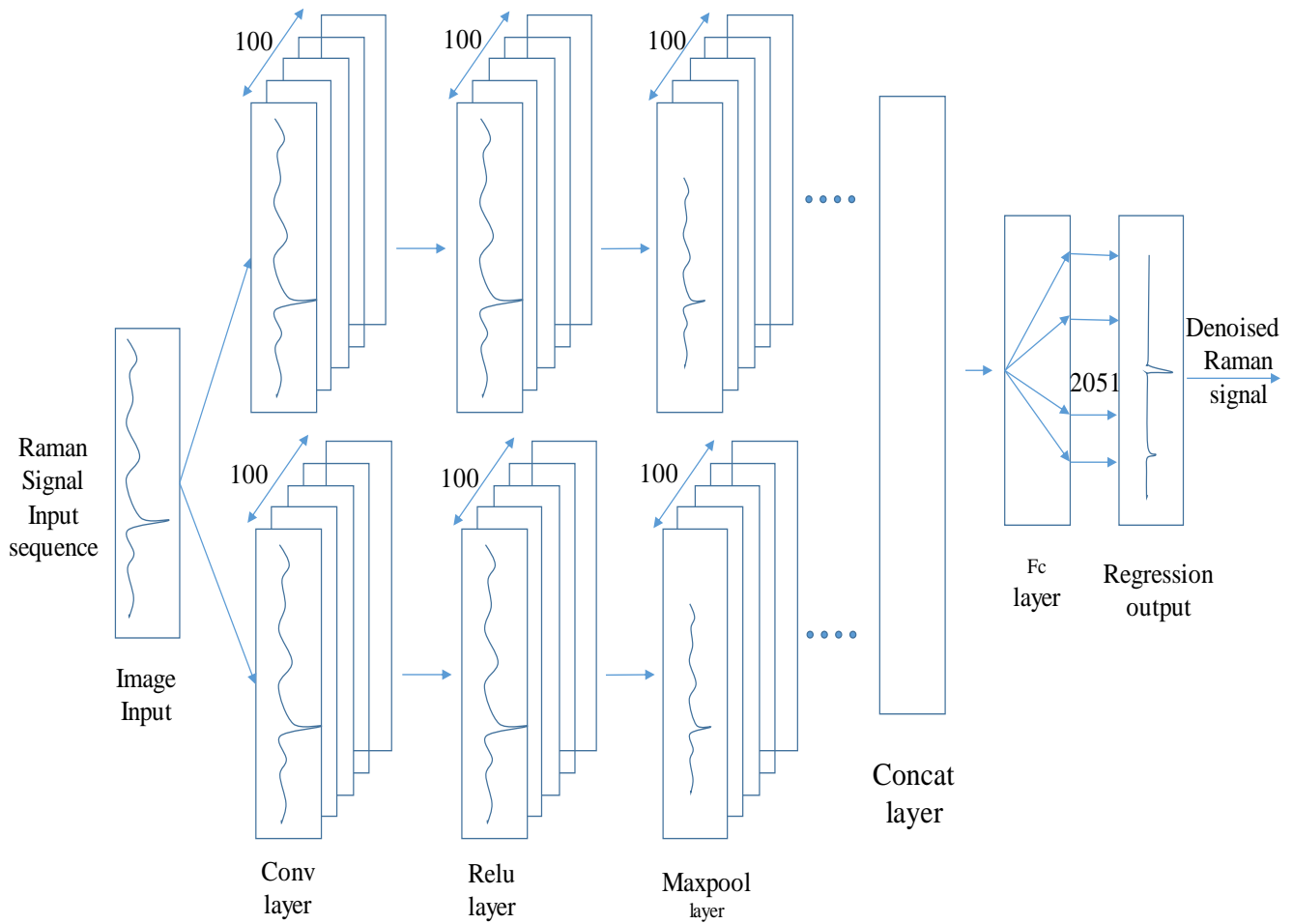

Fig. 3 Proposed structure of DL network.

Raman spectrum signal. The epoch passes through the entire dataset, while iteration is the mini-batch data for computing gradient and network parameters. An epoch goes through the entire dataset while an iteration is the calculation of the gradient and the network parameters for the mini-batch data.

Since convolution layer and pooling layer are the key of this experiment, we only need to introduce how convolution layer and pooling layer work. As shown in Fig. 4, all the signals are arranged in the form of a matrix, which is 2,051 and 15,390 wide. The matrix is processed on the principle of first-in/first-out. When the convolution kernel of 1×1 starts to work, it will do matrix operation along the wide direction. Because the width of the convolution kernel is 1, only one signal can be read for processing at a time. This is equivalent to that each neuron is only connected with the corresponding signal features, which greatly reduces the number of weights. In the convolution process, the weight of the convolution kernel will not change, which means that the same target features at different positions of the same signal are basically the same[9]. Each convolution layer of DL model has 50 convolution kernels. Each convolution kernel learns different features to extract Raman spectrum signal features. In this layer, 100 channels transfer the feature matrix to the pooling layer efficiently.

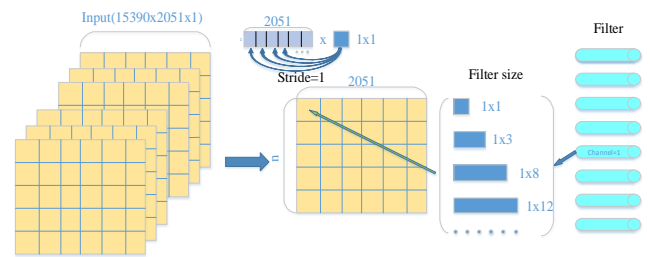

Fig. 4 Working principle diagram of the convolution layer.

The function of the pooling layer is similar to that of the convolution layer, but the signal characteristic matrix will be processed better in the downsampling process[10]. As shown in Fig. 5, the size of the core determines the size of each processing data. Its main function is downsampling, which makes some error values directly filtered in the process, leaving only excellent signal characteristics. Since the feature matrix will become smaller and smaller in this step, the signal features will gradually reduce the dimension in the neural network. To avoid these features disappear with the deepening of the DNN, we add 0 around the downsampling matrix to keep the same size as the convoluted feature matrix.

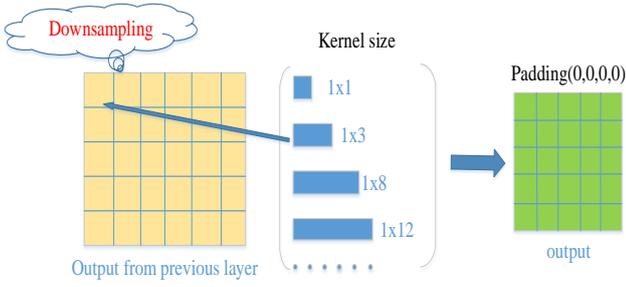

Fig. 5 Working principle diagram of pool layer.

After the multi-layer CNN, the Raman spectrum characteristic matrix will be expanded in the fully connected layer and output by the regression layer. We use the mean square error as the loss function. It can be seen as the result of the least-squares method compared to the signal length, and its core idea is the sum of squared errors of the true value ($y_i$) and the predicted value ($\hat{y}_i$). The calculation formula is defined as:

$$E = \sum_{i=0}^{n}(y_i - \hat{y}_i)^2 \quad (1)$$

To train the two-dimensional (2D) DL model, Adam optimization algorithm is selected as the global optimizer. The initial learning rate is set to $10^{-4}$. During the training process, the initial ratio decreased. The 2D DL model is trained on NVIDIA RTX 2080s. A total of 600 epochs are trained, and each epoch processes 50 signals. Using our computing resources, the training time of the model is about 120 minutes and the average test time is less than one millisecond.

Another noise suppression model is wavelet decomposition. As a high-performance signal processing tool, the wavelet transform is widely used in signal processing. Compared with Fourier transform and window Fourier transform, wavelet transform is a local transform of space (time) and frequency. It can analyze functions or signals in multi-scale or multi-resolution through stretching and translation operations, so it can extract information from signals more effectively. In contrast, the wavelet transform of input signal $x(t)$ is as follows:

$$W_{a,b} = \int_{-\infty}^{\infty} x(t) \frac{1}{\sqrt{a}} \gamma^*\left(\frac{t-b}{a}\right) dt \quad (2),$$

where $W_{a,b}$ is the wavelet transformation of $x(t)$, $a$ is the dilation parameter, $b$ is the location parameter and $\gamma^*(t)$ is the complex conjugate of the wavelet function.

Wavelet method is based on empirical Bayes, block James Stein, false discovery rate, minimax estimation, Stein's unbiased risk estimate, universal threshold g method to denoise Raman spectrum. It is implemented in MATLAB™ for denoising. The default is sym4 wavelet, where 4 is the number of lost moments. The other choices of this function include the selection of wavelet, threshold rule and the method of estimating noise variance. There are many methods based on wavelet denoising, and the experiment also tries to choose various parameters. Therefore, there are many improved methods based on wavelet denoising. To compare with the DL network denoising method, this experiment will be based on wavelet and use different methods for noise reduction comparison.

## B. Evaluation Index

In this paper, the output signal-to-noise ratio (SNR), root mean square error (RMSE) and mean absolute percentage error (MAPE) are selected as the evaluation indexes of denoised Raman spectrum signal. These indexes have the following characteristics: ① the noise is inversely proportional to the output SNR. When the output SNR is higher, the noise of Raman spectrum signal is smaller. ② The RMSE and the MAPE are proportional to the signal quality. When the RMSE and the MAPE are large, noise in the Raman spectrum signal is large, and the similarity between the original signal and the noise Raman spectrum signal is lower.

1.SNR

The experiment first uses the input and output SNR to evaluate the magnitude of the output noise. $P_s$ is the signal power and $P_n$ is the noise power. SNR is defined as:

$$\text{SNR} = 10\log_{10}\left(\frac{P_s}{P_n}\right) \quad (3)$$

2.RMSE

RMSE calculates the square root of the ratio of the square of the deviation between the predicted value and the true value and the number of signal samples *n*. RMSE is defined by,

$$RMSE = \sqrt{\frac{\sum_{t=1}^{n}|original-forecast|^2}{n}} \quad (4),$$

where *original* is the noiseless signal and *forecast* represents the denoised signal.

3.MAPE

MAPE measures the percentage of the predicted result to the true result which is defined by.

$$MAPE = \frac{1}{n}\sum_{t=1}^{n}\left|\frac{original-forecast}{orginal}\right| * 100\% \quad (5),$$

where *original* is the noiseless and *forecast* represents the recovered Raman signals.

## III. RESULTS AND DISCUSSION

In this section, we first visualize the features learned by DL model and the signal feature map shows the actual effect of DL network noise reduction. Secondly, in the case of the same noise Raman spectrum signal as the input, the noise reduction effect of DL network and wavelet based methods are compared and the evaluation index is used to express. Finally, the original signal and the denoised Raman signal are put together to intuitively compare the similarity between the denoised signal and the original signal, and generate a DL network noise reduction model.

### A. Experimental Results

Due to the number and size of convolution kernels in convolution layer, the pooling method of pooling layer will affect the efficiency of regression network. After the Raman spectrum signal is fed to the convolution and pooling layers, the DL network will extract part of the information of the noise Raman spectrum signal, and then reduce a part of the noise. As shown in Fig. 5 and 6, the noise signal passes through the convolution and the pooling layers. Because different parameters will affect the effect of noise reduction, in most cases, the feature map after pooling layer shows low noise Raman spectrum signal. In Fig. 6, noise level of some noisy signals does not decrease much after passing through the convolution layer, but some signals lose their original characteristics. This shows that the function of the convolution layer has no high noise reduction effect.

In Fig. 7, we can find that the pooling layer will reprocess and de-noising the signals from the convolution layer. In some channels of the convolution layer, the noisy Raman spectra have well displayed the original spectral characteristics.

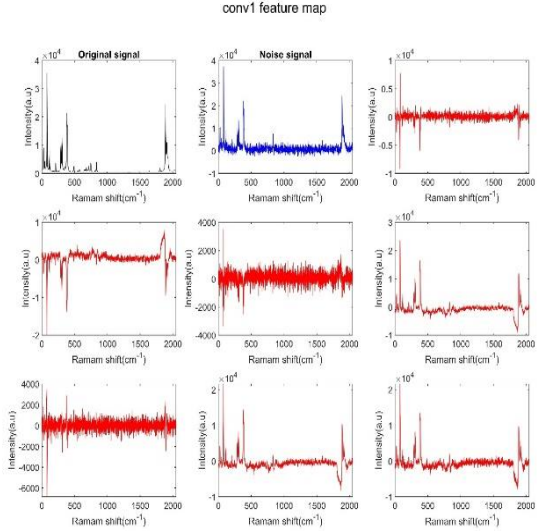

Fig. 6 Raman spectrum feature map after passing through the convolutional layer.

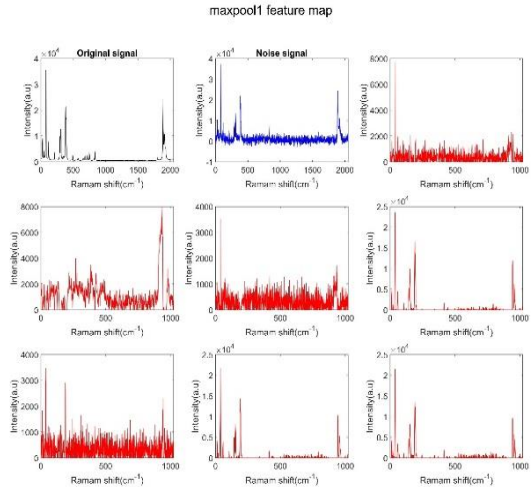

Fig. 7 Raman spectrum feature map after passing through the pooling layer.

For the Raman spectrum signal denoised by DL network, we use the output SNR, RMSE, MAPE to measure. In Table. 2, the indexes for evaluating the noise reduction effect of different methods are summarized. The input SNR of noisy Raman spectrum signal is 9.5 dB. After noise reduction by the DL network, the SNR signal is increased to 26.50 dB, which is much better than the original SNR. From the analysis of RMSE and MAPE, the difference between the denoised Raman spectrum signal and the noiseless Raman spectrum signal is small. Compared with the series CNN network, the parallel DL network has a greater noise reduction advantage. Based on the wavelet denoising platform, the empirical Bayes method has the best denoising performance and the Universal Threshold G method has the worst performance. Because RMSE and MAPE are also excellent indicators to measure different noise reduction methods. Through these two indicators, it is found that the Empirical Bayes noise reduction method is better than other methods. Comparing with the DL model, the CNN model and empirical Bayes noise reduction method, the DL model has obvious advantages in noise reduction.

TABLE I. Statistics of noise reduction results of DL, CNN and wavelet-based methods.

| Raman spectrum | | SNR (dB) | RMSE | MAPE |
|---|---|---|---|---|
| Noisy signal | | 9.50 | 1097.10 | 114.90 |
| DL | | 26.50 | 129.96 | 9.50 |
| CNN | | 23.49 | 183.75 | 13.79 |
| Wavelet noise reduction | Empirical Bayes | 16.25 | 422.59 | 27.07 |
| | Block James-Stein | 14.98 | 489.30 | 27.57 |
| | False Discovery Rate | 15.38 | 467.36 | 27.79 |
| | Minimax Estimation | 14.57 | 512.98 | 28.15 |
| | Stein's Unbiased Risk Estimate | 16.38 | 416.64 | 34.80 |
| | Universal Threshold G | 12.19 | 674.92 | 29.48 |

*B. Discussion*

The performances of wavelet methods for denoising are compared with DL based denoising. Because wavelet denoising method has a high position in the processing of noise signal, and because of the good learning ability of DL network, it can learn the signal and noise separately. To realize the noise reduction in the noise signal. Wavelet denoising method is better than DL network in terms of output SNR, RMSE and MAPE. As shown in Fig. 8, the original Raman spectrum signal, the signal after DL network denoising method is compared with the signal with the best denoising effect of wavelet. However, we find that DL network has better noise reduction effect than other methods. Different sizes of noise are reduced or eliminated almost entirely. We found that the signal after wavelet denoising still has some noise, some noise interferes with peak information, and some noise affects the judgment of sub-peak value.

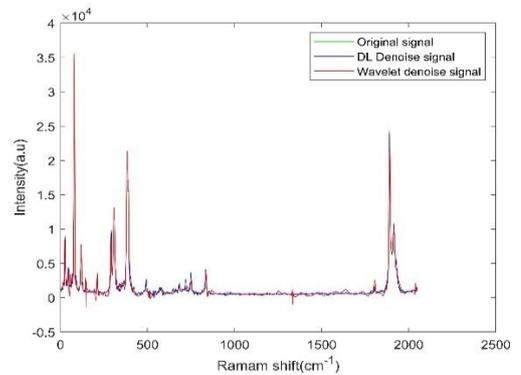

Fig. 8 Comparison of denoising performance between the DL and wavelet denoising based methods.

## IV. Conclusion

In this paper, we propose a DL model for noise reduction of Raman spectra. Since the absolute noise can not be avoided in the process of Raman spectrum extraction, wavelet denoising method can effectively reduce the noise of Raman spectrum signal. However, in analyzing the three evaluation indicators of noise reduction performance, we found that the output SNR based on DL network noise reduction technology is 10.24 dB larger than that of wavelet noise reduction method, RMSE is smaller than wavelet noise reduction method 292.63, and MAPE is smaller than wavelet Noise reduction method 10.09. This shows that the DL network denoising method has completely surpassed the wavelet denoising method in noise reduction. Although the denoising process of the two methods is different, it is found that the Raman spectrum signal processed by the DL network denoising method has a high similarity with the original Raman spectrum signal, and there are some differences in some peak values between the Raman spectrum signal processed by the wavelet denoising method and the original Raman spectrum signal.

## Acknowledgment

Thank to the chemical laboratory, Faculty of Science, Prince of Songkla University for the Raman spectral signal of the chemical substance.